\documentclass[a4paper,11pt]{article}
\usepackage[toc,page]{appendix}
\usepackage{graphicx}
\usepackage{colortbl}
\usepackage{amsmath}
\usepackage{subfigure}
\usepackage{mathtools}
\usepackage[utf8]{inputenc}
\usepackage{float}
\usepackage[normalem]{ulem}
\usepackage{epstopdf}
\usepackage{amsfonts,amsmath,amssymb}
\usepackage{hyperref}

\usepackage{epsfig,multicol,bbm}
\usepackage{color}
\usepackage[dvipsnames]{xcolor}

\definecolor{darkblue}{rgb}{0.0, 0.0, 0.55}
\definecolor{grey}{rgb}{0.57, 0.64, 0.69}
\definecolor{lightbrown}{rgb}{0.71, 0.4, 0.11}

\newcommand{\tcr}{\textcolor{red}}

\newcommand{\be}{\begin{equation}}
\newcommand{\ee}{\end{equation}}

\newcommand\fverb{\setbox\pippobox=\hbox\bgroup\verb}
\newcommand\fverbit{\egroup\item[\fbox{\unhbox\pippobox}]}

\newbox\pippobox

\begin{document}
\title{\bf Hairy Black holes in General Minimal Massive Gravity}
\author{ M. R. Setare\thanks{Electronic address: rezakord@ipm.ir},\,Julio Oliva\thanks{Electronic address: julioolivazapata@gmail.com},\,S. N. Sajadi \thanks{Electronic address: naseh.sajadi@gmail.com}
\\
\small Department of Science, Campus of Bijar, University of Kurdistan, Bijar, Iran\\
\small Departamento de F\'{\i}sica, Universidad de Concepci\'{o}n, Casilla,160-C, Concepci\'{o}n, Chile. 
}

\maketitle
\begin{abstract}
In this work, we investigate the near horizon and asymptotic symmetries of static and rotating hairy$-$AdS black hole in the framework of general minimal massive gravity. We obtain energy, angular momentum and entropy of the solutions.  Then we show that our results for these quantities are consistent with the first law of black hole thermodynamics. 
By considering the near horizon geometry of black hole, we find near horizon conserved charges and their algebra. By writing the algebra of conserved charges in terms of Fourier modes we have obtained the conserved charges in terms of zero modes.
\end{abstract}

\maketitle
\section{Introduction}
General Minimal Massive Gravity (GMMG) was introduced in \cite{Setare:2014zea}, providing a new example of a theory that avoids the bulk$-$boundary clash and therefore, as Minimal Massive Gravity (MMG) \cite{Bergshoeff:2014pca}, the theory possesses both, positive energy excitations around the maximally AdS$_3$ vacuum as well as a positive central charge in the dual CFT. Such clash is present in the previously constructed gravity theories with local degrees of freedom in 2+1$-$dimensions, namely Topologically Massive Gravity \cite{Deser:1982vy}, \cite{Deser:1981wh} and the cosmological extension of New Massive Gravity \cite{Bergshoeff:2009hq}. As described in the next section, the action principle of GMMG makes use of two auxiliary one$-$forms, $h$ and $f$, which at the level of the field equations can be integrated out, leading to the equations for New Massive Gravity supplemented by the Cotton tensor as well as by a parity even tensors, $J_{\mu\nu}$. The latter is quadratic in the curvature, and therefore the field equations for the metric remain of fourth order (see eq(13) of \cite{Setare:2014zea}). These effective Einstein equations cannot be obtained from a variational principle containing only the metric as a dynamical field, nevertheless they are on$-$shell consistent as it is the case in MMG \cite{Bergshoeff:2014pca}, as well as in the theories more recently introduced in \cite{Tekin:2015rha}, \cite{Ozkan:2018cxj} and \cite{Alkac:2018eck}.

Since these theories avoid the bulk-boundary clash, they define excellent arenas to explore the structure of asymptotically AdS solutions, asymptotic symmetries, their algebra and other holographically inspired questions. In this paper we explore many of these ideas in the context of GMMG.

In Section 2, we present the theory and show that it admits new static black hole solutions which are characterized by two integration constants and that were originally found as solutions of conformal gravity in 2+1 dimensions \cite{Oliva:2009hz} as well as on NMG \cite{Oliva:2009ip}, \cite{Bergshoeff:2009aq}. We also review the computation of the charges in Chern$-$Simons$-$like theories and obtain the thermodynamics quantities of the mentioned black hole solution of GMMG. We embed as well these solutions in a family of relaxed asymptotic behavior, and show that also in the context of GMMG, the charges associated to the extra U(1) current vanish. Then we turn our attention to the near horizon asymptotic symmetries and obtain the same structure that was recently obtained in the context of NMG in \cite{Donnay:2020yxw}. Section 3 is devoted to the extension of the previous analysis to the case of the rotating black hole, which can also be embedded in GMMG. We provide some conclusions in Section 4.

\section{Static Solution}\label{sec2}
The Lagrangian of GMMG is obtained by generalization the Lagrangian of generalized massive gravity.
The Lagrangian of GMMG model is
\cite{Setare:2014zea}
\begin{small}
\begin{equation}\label{eqlag}
L_{GMMG}=-\sigma e.R+\dfrac{\Lambda_{0}}{6}e.e\times e+h.T(\omega)+\dfrac{1}{2\mu}\left(\omega.d\omega+\dfrac{1}{3}\omega.\omega\times \omega \right)-\dfrac{1}{m^2}\left(f.R+\dfrac{1}{2}e.f\times f\right)+\dfrac{\alpha}{2}e.h\times h \ .
\end{equation}
\end{small}
Here $m$ is the mass parameter of NMG term, $h$ and $f$ are auxiliary one-form fields, 
${\Lambda}_{0}$ is a cosmological parameter with dimension of mass squared, $\sigma$ is a convenient sign, $\mu$ is a mass parameter of Lorentz Chern$-$Simons term, $\alpha$ is a dimensionless parameter, $e$ is a dreibein and $\omega$ is a dualized spin-connection. The equations of motion of the above Lagrangian by making variation with
respect to the fields $e$, $\Omega=\omega-\alpha h$, $h$ and $f$ and setting $\Lambda=-l^{-2}$ are as follows \cite{Setare:2017mry} (see appendix \ref{app0})
 \begin{align}\label{eqcc}
& \dfrac{\sigma}{l^2}-\alpha(1+\alpha\sigma)c_{h}^2+\Lambda_{0}-\dfrac{c_{f}^{2}}{m^2}=0,\\
 &\dfrac{1}{\mu l^2}-2(1+\alpha\sigma)c_{h}-\dfrac{2\alpha}{m^2}c_{f}c_{h}-\dfrac{\alpha^2}{\mu}c_{h}^2=0,\\
 &c_{f}-\mu(1+\alpha\sigma)c_{h}-\dfrac{\mu\alpha}{m^2}c_{f}c_{h}=0 .
 \end{align}
 % \begin{align}\label{eqcc}
%& 2\mu c_h (1+\sigma \alpha)+\Lambda +\alpha^2 c_h^2 +\dfrac{2\mu \alpha}{m^2}c_h c_f=0,\nonumber\\
% &\Lambda+\alpha^2 c_h^2+2c_f=0,\nonumber\\
 %&\sigma \Lambda -\Lambda_0+\alpha (1+\sigma \alpha)c_h^2+\dfrac{c_f^2}{m^2}=0\ .
 %\end{align}
By solving above equations, one can obtain $\Lambda_{0}$, $c_{h}$ and $c_{f}$ in terms of the couplings $\sigma,\mu,\alpha$ and $m^2$ as follows: 
 \begin{equation}\label{eqqcf}
c_{f}=-\dfrac{3m^{2}}{\alpha}+\dfrac{\sqrt{3}m}{\mu \alpha \mathcal{C}^{\frac{1}{6}}}\left(\sqrt{\mathcal{E}}\pm\dfrac{\sqrt{\mathcal{D}}}{\mathcal{E}^{\frac{1}{4}}}\right),
\end{equation}
\begin{equation}\label{eqqch}
c_{h}=\dfrac{m^{2}\left(-3m\mu \mathcal{C}^{\frac{1}{6}}\mathcal{E}^{\frac{1}{4}}+\sqrt{3}\mathcal{E}^{\frac{3}{4}}\pm\sqrt{3\mathcal{D}}\right)}{\mu \alpha\left(6\mu \alpha m \sigma \mathcal{C}^{\frac{1}{6}}\mathcal{E}^{\frac{1}{4}}+3m \mu \mathcal{C}^{\frac{1}{6}}\mathcal{E}^{\frac{1}{4}}+\sqrt{3}\mathcal{E}^{\frac{3}{4}}\pm\sqrt{3\mathcal{D}}\right)},
\end{equation}
which explicit form of $\mathcal{C}, \mathcal{D}$ and $\mathcal{E}$ have been provided in appendix \ref{app0}. 
For the typical values of parameters $\sigma=\alpha=m=\mu=1,\Lambda=-0.1$, we have $c_{f}=1.82,c_{h}=0.48, \Lambda_{0}=-3.9$. This shows that there are solutions with real $c_f$, $c_h$ and negative $\Lambda_{0}$. 
{Denoting the vacuum solution, the mass of gravitons obtained as (see appendix \ref{app01})
\begin{equation}
\Xi_{\pm}=-\dfrac{s {m}^2}{2{\mu}}\pm\sqrt{\dfrac{1}{2l^2}+\bar{\sigma}^2s {m}^{2}+\dfrac{{m}^4}{4{\mu}^2}-\dfrac{\bar{\sigma} s m^2 \gamma}{2\mu^2 l^2}}.
\end{equation}
 In the case of $\mu\to \infty$, the mass becomes $\Xi_{\pm}=\sqrt{1/2l^2+s\sigma^2m^2 }$ \cite{Bergshoeff:2009aq}. In the case of flat space limit $l\to \infty$, GMMG describes doublet massive excitation with mass $\Xi_{\pm}=-\frac{s {m}^2}{2{\mu}}\pm\sqrt{\bar{\sigma}^2s {m}^{2}+\frac{{m}^4}{4{\mu}^2}}$.
In order to have a positive mass we should have ($\Xi_{\pm}>0$):
\begin{equation}
 (m^2 l)^2(s^2-1)<2\mu^2 +4 s m^2\bar{\sigma}^2 \mu^2 l^2-2s\bar{\sigma} m^2\gamma \mu^2.
\end{equation}}
Once the auxiliary one$-$form fields $f$ and $h$ are integrated out, the field equations can be written as
\begin{equation}\label{eqfield}
\bar{\sigma}G_{\mu\nu}+\bar{\Lambda}_0g_{\mu\nu}+\frac{1}{\mu}C_{\mu\nu}+\frac{\gamma}{\mu^2}J_{\mu\nu}+\frac{s}{2m^2}K_{\mu\nu}\ ,
\end{equation}
where $C_{\mu\nu}$ is the Cotton tensor, $K_{\mu\nu}$ is the Euler$-$Lagrange derivative of the quadratic part of the NMG Lagrangian with respect to the metric, and $J_{\mu\nu}$ is the quadratic in the curvature tensor introduced in \cite{Bergshoeff:2014pca}. The parameter $s$ is sign, $\gamma$, $\bar{\sigma}$ and $\bar{\Lambda}_{0}$ are the parameters which defined in terms of other parameters like $\sigma, m$ and $\mu$.

The field equations (\ref{eqfield}) admits a 1$-$parameter hairy generalization of BTZ solution. In the static case, the metric of hairy$-$AdS black hole takes the form \cite{Oliva:2009ip}, \cite{Bergshoeff:2009aq}
\begin{equation}\label{eqmetric}
ds^{2}=-\left(\dfrac{r^2}{l^2}+b r-\mathcal{M}\right)dt^{2}+\left(\dfrac{r^2}{l^2}+b r-\mathcal{M}\right)^{-1}dr^{2}+r^{2}d\phi^{2}
\end{equation}
where $t\in R$, $\phi \in [0,2\pi]$, $r\in R_{>0}$  and $\mathcal{M}$, $b$ are two real integration constants. For $b = 0$ the metric \eqref{eqmetric} reduces to the BTZ black hole. Defining $\Lambda=-l^{-2}$, the metric (\ref{eqmetric}) is a solution to the GMMG field equations provided
\begin{equation}\label{eqsp}
\bar{\Lambda}_{0}=-\dfrac{\bar{\sigma}}{2l^{2}},\hspace{1cm}m^{2}=\dfrac{\mu^{2}s}{\gamma +2 \bar{\sigma} l^2 \mu^2},
\end{equation}
where $\gamma=-2\alpha c_{h}\mu l^2/3$, $\bar{\sigma}=\sigma$, $s=-2c_{f}l^2$. So, the metric (\ref{eqmetric}) with conditions (\ref{eqsp}), is a solution for field equation (\ref{eqfield}).
The Cotton tensor of the spacetime \eqref{eqmetric} vanishes since it is conformally flat. This is consistent the fact that the TMG mass parameter in Equations \eqref{eqsp}, appear only through the combination $\gamma/\mu^2$, namely the coupling of the tensor $J_{\mu\nu}$, which is generically non$-$vanishing on this spacetime. Equations \eqref{eqsp} relate the couplings in a manner that is different than that which leads to the chiral point of GMMG \cite{Setare:2014zea} (see also \cite{Giribet:2014wla} for the definition of Log Minimal gravity arising at the chiral point of MMG).
For a certain range of the parameters $\mathcal{M}$ and $b$, the solution describes a black hole with the Killing horizons located at $r_{\pm}$ such that $f(r_{\pm})=0$, namely
\begin{equation}
r_{\pm}=\dfrac{1}{2}\left(-b l^{2}\pm l \sqrt{b^2 l^2+4\mathcal{M}}\right),
\end{equation}
which leads also to
\begin{equation}
b=-\dfrac{(r_{+}+r_{-})}{l^2},\hspace{1cm}\mathcal{M}=-\dfrac{r_{+}r_{-}}{l^2}.
\end{equation}
 Here, using an extension of the ADT method for Chern$-$Simons$-$like theories of gravity \cite{Bergshoeff:2019rdb}, we will obtain the conserved charges that characterize, globally, the black hole solution \eqref{eqmetric} of GMMG. 
Now, considering the Killing vector $\partial_t$ of the black hole vacuum background \eqref{eqmetric}, we find that (see appendix \ref{app02})
\begin{small}
\begin{align}\label{eqmasssta1}
&E=Q(\partial_t)=\dfrac{1}{8G}\left(\sigma+\dfrac{c_f}{m^2}\right)(4\mathcal{M}+b^2 l^2)=\dfrac{1}{8G l}\left(\sigma+\dfrac{c_f}{m^2}\right)(r_{+}-r_{-})^2\ .
\end{align}
\end{small}
As in NMG \cite{Donnay:2020yxw}, the extremal hairy black hole $r_{+}=r_{-}$, is also massless in GMMG. The Hawking temperature of the black hole reads
\begin{equation}\label{tem}
T=\dfrac{\kappa}{2\pi}=\dfrac{\sqrt{b^2 l^2+4\mathcal{M}}}{4\pi l}=\dfrac{r_{+}-r_{-}}{4\pi l^2}.
\end{equation}
While the black hole entropy can be calculated from the Cardy formula proposed in \cite{Setare:2018lnf} and leads to
\begin{equation}\label{eqcardyen}
S=\dfrac{\pi^2 l}{3}(c_R T_R+c_L T_L)=\dfrac{\pi}{4G}\left(\bar{\sigma}+\dfrac{\alpha c_h}{\mu}+\dfrac{c_f}{m^2}\right)\left(r_{+}-r_{-}\right),
\end{equation}
where 
\begin{equation}\label{eqcentr}
c_{R/L}=\dfrac{3l}{2G}\left(\bar{\sigma}+\dfrac{\alpha c_h}{\mu}+\dfrac{c_f}{m^2}\mp \dfrac{l}{\mu }\right),\;\;\;\;\;T_{R/L}=\dfrac{\sqrt{4\mathcal{M}+b^2 l^2}}{4\pi l}.
\end{equation}
By considering the conserved charge $E$ as the thermodynamic energy, one can verify the first law of the black hole thermodynamics and Smarr$-$like relation as follow \cite{Blagojevic:2015zma}:
\begin{equation}
dE=TdS, \hspace{0.5cm} E=\dfrac{1}{2} T S.
\end{equation}
In the case of $\alpha=0$, the theory goes to GMG theory \cite{Bergshoeff:2019rdb}. In this case, the relations \eqref{eqqcf} and \eqref{eqqch} become
\begin{equation}\label{eqqchcf}
c_{h}=\dfrac{1}{2\mu l^2},\;\;\;\;\;\; c_{f}=\dfrac{1}{2l^2},\;\;\;\;\;\;\Lambda_{0}=-\dfrac{\sigma}{l^2}+\dfrac{1}{4m^2 l^2}.
\end{equation} 
For the static BTZ black hole with $r_{+}=-r_{-}$ in \eqref{eqmasssta1}-\eqref{eqcentr}, the thermodynamical quantities become \cite{Bergshoeff:2019rdb}
\begin{equation}\label{eqqes}
E=\dfrac{1}{2G l}\left(\sigma+\dfrac{1}{2m^2 l^2}\right)r_{+}^2,\;\;\;\; S=\dfrac{\pi}{2G}\left(\bar{\sigma}+\dfrac{1}{2m^2 l^2}\right)r_{+}.
\end{equation} 
In the case of $\alpha=0$ and $m\to \infty$, the results \eqref{eqqchcf} and \eqref{eqqes} go to the case of BTZ in TMG \cite{Bergshoeff:2019rdb}.

\subsection{ASYMPTOTIC SYMMETRIES}
The presence of the linear term in the radial coordinate, in the lapse function of the black hole \eqref{eqmetric}, forces us to consider asymptotic conditions in GMMG which are relaxed with respect to the Brown$-$Henneaux asymptotic behavior of General Relativity \cite{Brown:1986nw}. Writting the metric as $g_{\mu\nu}=\bar{g}_{\mu\nu}+h_{\mu\nu}$ we consider the following set of asymptotic conditions, which accommodate the hairy black hole \eqref{eqmetric}
\begin{align}\label{eqcond}
 h_{rr}=&h^{'}_{rr} r^{-3}+f_{rr}r^{-4}+\cdots
 \nonumber\\
 h_{r i}=&h^{'}_{r i}r^{-1}+f_{r i}r^{-2}+\cdots\nonumber\\
 h_{i j}=&h^{'}_{i j}r+f_{i j}+\cdots \ , 
\end{align}
where we will consider the AdS$_3$ background in null coordinates $x^{\pm}=t\pm \phi$  and $r$ as
\begin{equation}
d\bar{s}^2=-l^2 (dx^{+})^2-l^2 (dx^{-})^2-2(l^2+2r^2)dx^{+}dx^{-}+\dfrac{l^2 dr^2}{l^2+r^2}\ .
\end{equation}
The Brown$-$Henneaux asymptotic behavior is controlled by the $f_{\mu\nu}$ terms in \eqref{eqcond}, while the $h'_{\mu\nu}$ terms represent a weakening with respect to the Brown$-$Henneaux behaviour, which are necessary if one aims to include the black hole \eqref{eqmetric} as a physically relevant solution of GMMG.

The asymptotic behavior \eqref{eqcond} is preserved by the action generated by the Lie derivative along the following asymptotic Killing vectors
\begin{align}
\xi^{r}  & =-\frac{r}{2}\left(  \partial_{+}L^{+}+\partial_{-}L^{-}\right)
+...\\
\xi^{+}  & =L^{+}+\frac{l^{2}}{2r^{2}}\partial_{-}^{2}L^{-}+...\\
\xi^{-}  & =L^{-}+\frac{l^{2}}{2r^{2}}\partial_{+}^{2}L^{+}+...
\end{align}
where $L^+=L^+\left(x^+\right)$ and $L^-=L^-\left(x^-\right)$, and $(...)$ stands for subleading terms in the $r\rightarrow \infty$ expansion.

It was noticed in \cite{Giribet:2009qz,Donnay:2020yxw}, that the vector field
\begin{equation}\label{eqzeta}
\zeta =Y(x^{+},x^{-})\partial_{r}
\end{equation}
also preserves the boundary conditions (\ref{eqcond}). Since
\begin{align}
\mathcal{L}_{\zeta}g_{r r}&=-\dfrac{2l^2 Y r}{(r^2+l^2)^2}=\mathcal{O}\left(r^{-3}\right),\;\;\; \mathcal{L}_{\zeta}g_{- r}=\dfrac{l^2 \partial_{-}Y}{r^2+l^2}=\mathcal{O}\left(r^{-2}\right),\;\;\;\mathcal{L}_{\zeta}g_{+ r}=\dfrac{l^2 \partial_{+}Y}{r^2+l^2}=\mathcal{O}\left(r^{-2}\right)\nonumber\\
\mathcal{L}_{\zeta}g_{+ +}&=\mathcal{L}_{\zeta}g_{- -}=0=\mathcal{O}\left(1\right),\;\;\;\mathcal{L}_{\zeta}g_{+ -}=-4 Y r=\mathcal{O}(r)\ .
\end{align}
Consequently, the asymptotic Killing vectors $\xi$ and $\zeta$ generate two copies of the Witt algebra\\
\begin{equation}
\left[\eta(L^{+}_{1},L^{-}_{1}),\eta(L^{+}_{2},L^{-}_{2})\right]=\hat{\eta}(L^{+},L^{-})
\end{equation}
with 
\begin{equation}
\hat{L}^{+}=L_{1}^{+}\partial_{+}L^{+}_{2}-L^{+}_{2}\partial_{+}L^{+}_{1},\hspace{1cm}\hat{L}^{-}=L_{1}^{-}\partial_{-}L^{-}_{2}-L^{-}_{2}\partial_{-}L^{-}_{1}
\end{equation}
in semi$-$direct sum with an infinite$-$dimensional Abelian ideal:
\begin{equation}
\left[\zeta(Y_{1}),\zeta(Y_{2})\right]=0,\hspace{1cm}\left[\zeta(Y_{1}),\eta(L^{+}_{2},L^{-}_{2})\right]=\zeta(\hat{Y})
\end{equation}
with
\begin{equation}
\hat{Y}=-\dfrac{1}{2}Y_{1}(\partial_{-}L^{-}_{2}+\partial_{+}L^{+}_{2})-L^{-}_{2}\partial_{-}Y_{1}-L^{+}_{2}\partial_{+}Y_{1}
\end{equation}
in terms of the modified Lie brackets $\left[\zeta, \eta\right]=L_{\zeta}\eta-\delta_{\zeta}\eta+\delta_{\eta}\zeta$ introduced in \cite{Barnich:2010eb}.
The vector field $\zeta$ can be though of as an asymptotic supertranslation, and using \eqref{generalcharge} it can be shown to yield a vanishing charge in GMMG; namely
\begin{equation}
Q(\zeta)=0\ .
\end{equation}
This implies that supertranslation symmetry generated by (\ref{eqzeta}) is pure gauge also in the context of GMMG, as recently proved it is the case in NMG \cite{Donnay:2020yxw}.

Besides infinity, another special surface in a black hole spacetime is the event horizon, were based on previous experience one might find an enhanced group of asymptotic symmetries. The following section is devoted to such analysis in the context of the hairy black hole \eqref{eqmetric} of GMMG.
%-------------------
%-----Near Horizon
%-------------------
\subsection{NEAR HORIZON SYMMETRIES}
Adapting null Gaussian coordinates to an event horizon in $2+1$ dimensions (see e.g. \cite{Donnay:2015abr},\cite{Setare:2016jba}) allows writting the metric as
\begin{equation}
ds^2=f dv^2+2k dv d\rho +2h dv d\phi + R^2 d\phi^2
\end{equation}
where
\begin{align}\label{asympnh}
f=&-2\kappa \rho  +\tau(\phi)\rho^2+\mathcal{O}\left(\rho^3 \right)\nonumber\\
k=&1+\mathcal{O}\left(\rho^2 \right)\nonumber\\
h=&\theta(\phi)\rho +\sigma(\phi)\rho^2+\mathcal{O}\left(\rho^3 \right)\nonumber\\
R^2=&\Gamma^2(\phi)+\lambda (\phi)\rho +\mathcal{O}\left(\rho^2 \right)\  .
\end{align}
Here, the event horizon is located at $\rho=0$ and $\theta(\phi)$, $\Gamma(\phi)$ and $\lambda(\phi)$ are arbitrary functions of the coordinate $\phi$, while $\kappa$ which corresponds to the surface gravity is held constant. The dreibein can be chosen as
\begin{align}
e^{0}=\sqrt{-f+\dfrac{h^2}{R^2}}dv-\dfrac{k}{\sqrt{-f+\dfrac{h^2}{R^2}}}d\rho,\;\;\;\;\;\;
e^{1}=\dfrac{k}{\sqrt{-f+\dfrac{h^2}{R^2}}}d\rho,\;\;\;\;\;\;
e^{2}=\dfrac{h}{R}dv+Rd\phi.
\end{align}

The near horizon boundary conditions presented above are preserved by a set of asymptotically Killing vectors $\xi=\xi^{\mu}\partial_{\mu}$ which are \cite{Donnay:2015abr}
\begin{align}\label{horizonkilling}
\xi^{v}=&P(\phi)+...\nonumber\\
\xi^{\rho}=&\dfrac{\theta(\phi)}{2\Gamma^2(\phi)}\partial_{\phi}P(\phi)\rho^2 +...\nonumber\\
\xi^{\phi}=&L(\phi)-\dfrac{1}{\Gamma^{2}(\phi)}\partial_{\phi}P(\phi)\rho +\dfrac{\lambda(\phi)}{2\Gamma^{4}(\phi)}\partial_{\phi}P(\phi) \rho^2 +...
\end{align}
where again, the ellipsis denote subleading terms in the $\rho\rightarrow 0$ expansion. The conserved charges associated to these near horizon asymptotic Killing vectors, in the context of GMMG can be computed along the lines of \cite{Setare:2016jba}, leading to
\begin{equation}
Q(\xi)=\dfrac{1}{8\pi G}\int_{\rho\rightarrow 0} K(\xi)
\end{equation}
where
\begin{align}
K(\xi)=-\left(\sigma + \dfrac{\alpha c_h}{2\mu}+\dfrac{c_f}{m^2}\right)(i_{\xi}\Omega -\chi_{\xi}).e+\dfrac{1}{2\mu}(i_{\xi}\Omega -\chi_{\xi}).\Omega-\dfrac{\alpha c_h}{2\mu}i_{\xi}e.\Omega+\dfrac{1}{2\mu l^2}i_{\xi}e.e
\end{align}
and
\begin{equation}
(i_{\xi}\Omega -\chi_{\xi})^{a}=-\dfrac{1}{2}\epsilon^{a}{}_{b c}e^{b}_{\mu} e^{c}_{\nu}\nabla^{\mu}\xi^{\nu},\;\;\;\;\;\; \Omega^{\mu}_{\nu}=\dfrac{1}{2}\epsilon^{\mu \alpha \beta}e^{c}_{\beta}\nabla_{\nu}e_{c\alpha}.
\end{equation}
Using these expressions one explicitly obtains
\begin{align}\label{eqconsch}
Q(\xi)=&-\dfrac{1}{16\pi G}\int_{0}^{2\pi}d\phi \left[\left[\left(\bar{\sigma}+\dfrac{\alpha c_h}{2\mu}+\dfrac{c_f}{m^2}\right)\Gamma(\phi)+\dfrac{1}{4\mu}\theta(\phi)\right]\left[2\kappa P(\phi)-\theta(\phi)L(\phi)\right]\right.\nonumber\\
&\left.-\dfrac{\alpha c_h}{2\mu}\Gamma(\phi)\theta(\phi)L(\phi)-\dfrac{1}{\mu l^2}\Gamma(\phi)^2 L(\phi)\right]\ .
\end{align}
Notice that in opposition to the situation in NMG \cite{Donnay:2020yxw}, these charges do not depend on the subleading terms of the $g_{vv}$ component of the metric in \eqref{asympnh}. Comparing with the near horizon expansion of the solution (\ref{eqmetric}), we have \cite{Donnay:2020yxw} 
\begin{equation}\label{eqq39}
\kappa=\dfrac{(r_{+}-r_{-})}{2l^2},\hspace{0.5cm}\Gamma=r_{+},\hspace{0.5cm}\theta=0,\hspace{0.5cm} \lambda=2 r_{+},\hspace{0.5cm} \tau=-\dfrac{1}{l^2} .
\end{equation} 
In particular, the charges associated to the zero$-$modes of $P$ and $L$ lead to
\begin{equation}
Q=-\dfrac{r_{+}}{8G l^2}\left[\left(\bar{\sigma}+\dfrac{\alpha c_h}{2\mu}+\dfrac{c_f}{m^2}\right)(r_{+}-r_{-})P-\dfrac{r_+}{\mu}L\right].
\end{equation}
The algebra of conserved charges can be written as \cite{Barnich:2001jy}
\begin{equation}\label{eqcomut}
\left\lbrace Q(\xi_{1}),Q(\xi_{2})\right\rbrace=Q\left([\xi_{1},\xi_{2}]\right)+\mathcal{C}\left(\xi_{1},\xi_{2}\right)
\end{equation}
where $\mathcal{C}(\xi_{1},\xi_{2})$ is a central extension term. Also, the left hand side of the equation (\ref{eqcomut}) can be defined by \cite{Barnich:2001jy}
\begin{equation}
\left\lbrace Q(\xi_{1}),Q(\xi_{2})\right\rbrace= \dfrac{1}{2}\left(\delta_{\xi_{2}}^{(g)}Q(\xi_{1})-\delta_{\xi_{1}}^{(g)}Q(\xi_{2})\right)
\end{equation}
Therefore, using the conserved charges (\ref{eqconsch}), for the near horizons asymptotic Killing vectors, in GMMG we find that
\begin{equation}
\left\lbrace Q(\xi_{1}),Q(\xi_{2})\right\rbrace=Q\left([\xi_{1},\xi_{2}]\right)+\dfrac{\kappa}{64\pi \mu G}\int_{0}^{2\pi} d\phi \left[2\kappa (T_{1}T^{'}_{2}-T_{2}T_{1}^{'})-T_{12}(\theta(\phi)+2\alpha c_h\Gamma(\phi))\right].
\end{equation}
Introducing the charges associates to each Fourier mode $\mathcal{T}_{n}=Q\left(P(\phi)=e^{in\phi},L(\phi)=0\right)$ and $\mathcal{Y}_{n}=Q\left(P(\phi)=0,L(\phi)=e^{in\phi}\right)$, leads to
\begin{align}\label{furiealgebra}
&i\lbrace\mathcal{T}_{m},\mathcal{T}_{n}\rbrace=-\dfrac{\kappa^2 n}{8G\mu}\delta_{m+n,0},\hspace{0.5cm}i\lbrace\mathcal{Y}_{m},\mathcal{Y}_{n}\rbrace=(m-n)\mathcal{Y}_{m+n},\nonumber\\
&i\lbrace\mathcal{Y}_{m},\mathcal{T}_{n}\rbrace=-n\mathcal{T}_{m+n}+\dfrac{\kappa n}{64\pi G\mu}\int_{0}^{2\pi}e^{i(m+n)\phi}\left[\theta(\phi)+2\alpha c_h \Gamma(\phi)\right]d\phi.
\end{align}
And one can easily read off the eigenvalues of $\mathcal{T}_{n}$ and $\mathcal{Y}_{n}$ from (\ref{eqconsch})
\begin{equation}\label{eqqtau}
\mathcal{T}_{n}=-\dfrac{\kappa}{8\pi G}\int_{0}^{2\pi}e^{in\phi}\left[\left(\bar{\sigma}+\dfrac{\alpha c_h}{2\mu}+\dfrac{c_f}{m^2}\right)\Gamma(\phi)+\dfrac{1}{4\mu}\theta(\phi)\right]d\phi
\end{equation}
\begin{equation}\label{eqy}
\mathcal{Y}_{n}=\dfrac{1}{16\pi G}\int_{0}^{2\pi}e^{i n\phi}\left[\left(\bar{\sigma}+\dfrac{\alpha c_h}{\mu}+\dfrac{c_f}{m^2}\right)\Gamma(\phi)\theta(\phi)+\dfrac{\theta(\phi)^2}{4\mu}+\dfrac{\Gamma(\phi)^2}{\mu l^2}\right]d\phi.
\end{equation}
For the hairy black hole solution (\ref{eqmetric}), using (\ref{eqq39}), we have
\begin{equation}
\mathcal{T}_{n}=-\dfrac{1}{4G}\left(\bar{\sigma}+\dfrac{\alpha c_h}{2\mu}+\dfrac{c_f}{m^2}\right)\kappa r_{+}\delta_{n,0},
\end{equation}
\begin{equation}
\mathcal{Y}_{n}=\dfrac{r_{+}^2}{8G\mu l^2}\delta_{n,0}.
\end{equation}
We know that the zero mode charge $\mathcal{T}_{0}$ is proportional to the entropy of the black hole solution, i.e. $\mathcal{T}_{0}=\dfrac{\kappa}{2\pi}S$, where $S$ is the entropy of the black hole
\begin{equation}
S=-\dfrac{\pi}{2G} \left(\bar{\sigma}+\dfrac{\alpha c_h}{2\mu}+\dfrac{c_f}{m^2}\right) r_{+},
\end{equation}
and $\mathcal{Y}_{0}$ gives us the angular momentum of the black hole, i.e. $J=\mathcal{Y}_{0}$.
The algebra spanned by $\mathcal{T}_{n}$ and $\mathcal{Y}_{n}$ reduced to the following
\begin{align}
&i[\mathcal{T}_{m},\mathcal{T}_{n}]=-\dfrac{\kappa^2 n}{8\mu G}\delta_{m+n,0},\hspace{0.5cm}i[\mathcal{Y}_{m},\mathcal{Y}_{n}]=(m-n)\mathcal{Y}_{m+n},\nonumber\\
&i[\mathcal{Y}_{m},\mathcal{T}_{n}]=-n\mathcal{T}_{m+n}+\dfrac{\alpha c_h \kappa n r_{+}}{16\mu G}\delta_{m+n,0}.
\end{align}

In the covariant formalism, the functional variation of the conserved
charge associated to a given asymptotic Killing vector $\xi$ is given by the expression
\begin{equation}\label{eqcovp}
    \delta Q(\xi; g,\delta g)=\dfrac{1}{16\pi G}\int_{0}^{2\pi}d\phi \sqrt{-g}\epsilon_{\mu \nu \phi}k^{\mu \nu}_{\xi}
\end{equation}
where $g$ is a solution, $\delta g$ a perturbation around it, and $k^{\mu \nu}$ is a surface 1$-$form potential. Evaluating (\ref{eqcovp}) \cite{commingsoon},\cite{Devecioglu:2010sf}
\begin{equation}
    Q(\partial_{v})=\dfrac{\kappa}{16\pi G}\int_{0}^{2\pi}d\phi  \left[-\dfrac{\theta(\phi)}{\mu}+\dfrac{s} {4 m^2}\dfrac{\kappa \lambda(\phi)}{\Gamma(\phi)} \right]+D
\end{equation}
where $D$ stands for a non$-$integrable part that vanishes when 
$\gamma$, $\lambda$, $\theta$ and $\tau$ are constant.
The charge $Q$, associated to the zero$-$mode of supertranslation vector, in the case where 
$\Gamma, \theta$ and $ \lambda$ are independent of $\phi$,  is given by
\begin{equation}
 Q=\dfrac{\kappa}{8G}\left[-\dfrac{\theta}{\mu}+\dfrac{s} {4 m^2}\dfrac{\kappa \lambda(\phi)}{\Gamma(\phi)} \right].  
\end{equation}
By evaluating this charge for the hairy black hole geometry and using the relevant metric functions we have
\begin{equation}
Q=T\dfrac{\pi}{4G}\left[\dfrac{s}{m^2l^2}(r_{+}-r_{-})\right]=TS    \end{equation}
with
\begin{equation}
   S=\dfrac{\pi s}{4G m^2 l^2}(r_{+}-r_{-})=\dfrac{\pi}{4G}\left(2\sigma +\dfrac{\gamma}{\mu^2 l^2}\right) (r_{+}-r_{-})
\end{equation}
which is the same as equation (\ref{eqcardyen}), if 
\begin{equation}
    2\sigma +\dfrac{\gamma}{\mu^2 l^2}=\bar{\sigma}+\dfrac{\alpha c_h}{\mu}+\dfrac{c_f}{m^2}.
\end{equation}

\section{Rotating black hole}
A rotating generalization of the hairy black hole (\ref{eqmetric}) is given by
\begin{equation}\label{eqrot}
ds^2=-N^{2}(r)F(r)dt^{2}+\dfrac{dr^2}{F(r)}+(r^2+r_0^2)(N^{\phi}(r)dt+d\phi)^2
\end{equation}
where
\begin{align}
N^{2}(r)=&\dfrac{(4 r+b l^2(1-\eta))^2}{16 r^2+l^2(1-\eta)(8\mathcal{M}+b^2 l^2(1-\eta))}\nonumber\\
F(r)=&\dfrac{r^2}{l^2}+\dfrac{(\eta +1)b r}{2}-\mathcal{M}\eta + \dfrac{b^2 l^2 (1-\eta)^2}{16}\nonumber\\
N^{\phi}(r)=&\dfrac{8a(br-\mathcal{M})}{16 r^2+l^2(1-\eta)(8\mathcal{M}+b^2 l^2(1-\eta))}\nonumber\\
r_{0}^2=&\dfrac{l^2 (1-\eta)(8\mathcal{M}+b^2 l^2(1-\eta))}{16}
\end{align}
and $\eta=\sqrt{1-a^2/l^2}$ and $a$ is the rotation parameter, which was found as a solution of NMG in \cite{Oliva:2009ip}. When $a = 0$, the metric reduces to the static hairy black hole (\ref{eqmetric}), while for $b = 0$ it reduces to the stationary BTZ black hole. For certain range of the parameters $\mathcal{M}$ and $b$ the solution describes a black hole, with horizons located at ($F(r)=0$)
\begin{equation}
r_{\pm}=\dfrac{1}{2}\left(-\dfrac{l^2b(1+\eta)}{2}\pm l \sqrt{\eta(b^2 l^2+4\mathcal{M})}\right).
\end{equation}
 Provided the equations (\ref{eqsp}) are fulfilled, the metric (\ref{eqrot}) is a solution for GMMG.\\
 Following the method considered in Section II, we obtain the conserved charges of the rotating black hole. For the Killing vector $\partial_{t}$ one can obtain the mass as follows (see appendix \ref{app02})
\begin{align}\label{eqer1}
E=&Q(\partial_t)=\dfrac{1}{32 G}\left(\sigma+\dfrac{c_f}{m^2}\right)\left[4\mathcal{M}+b^2 l^2-\dfrac{b^2 a^2}{2}\right]+\dfrac{a}{32 G \mu l^2}\left[8\mathcal{M}+b^2 l^2 (1-\eta)\right],
\end{align} 
 and for Killing vector $\partial_\phi$, one can obtain angular momentum as follows 
\begin{align}\label{eqang1}
J=&Q(\partial_\phi)=\dfrac{1}{64 G}\left(\sigma+\dfrac{c_f}{m^2}\right)\left[8\mathcal{M}+b^2 l^2(1-\eta)\right]a+\dfrac{1}{32 G\mu}\left[(4\mathcal{M}+b^2 l^2) (1+\eta)-\dfrac{b^2 a^2}{2}\right]+\nonumber\\
&\dfrac{c_h l^2}{32 G}\left[(4\mathcal{M}+b^2 l^2)(1-\eta)-\dfrac{b^2 a^2}{2}\right].
\end{align} 
In the case of $\eta=1$, equation (\ref{eqer1}) reduces to equation (\ref{eqmasssta1}) for the static metric. But, for $b=0$ equation (\ref{eqer1}) does not reduce to BTZ black hole, because here we are in an unusual coordinate system. Also, notice that equation (\ref{eqang1}) in the case of $\eta=1$ does not equal to zero, as expected from other three dimensional gravitational theories containing parity$-$odd terms \cite{Carlip:1994hq}.
Angular velocity of outer horizon (inner horizon) is
\begin{equation}
\Omega_{\pm}=-N^{\phi}_{\pm}=\dfrac{-8a (b r_{\pm}-\mathcal{M}) }{16 r_{\pm}^2+(1-\eta)l^2(8 \mathcal{M}+b^2 +1)}=\dfrac{a}{(\eta +1)l^2}=\dfrac{1}{l}\sqrt{\dfrac{1-\eta}{1+\eta}},
\end{equation}
and $T_\pm$, the Hawking temperatures of the inner and outer horizons, are
\begin{equation}
T_{\pm}=\dfrac{\kappa_{\pm}}{2\pi}=\pm \dfrac{\sqrt{2(\eta +1)}}{8\pi l}\left(\sqrt{l^2 b^2+4\mathcal{M}}\right)=\pm \sqrt{\dfrac{2(\eta +1)}{\eta}}\dfrac{(r_{+}-r_{-})}{8\pi l^2}.
\end{equation}
 The black hole entropy from the Cardy formula can be calculated as \cite{Giribet:2009qz}
\begin{equation}\label{eqentropy}
S^{R/L}=\dfrac{\pi^2 l}{3}(c_{R}T_{R}\pm c_{L}T_{L})=\dfrac{\pi}{4}\dfrac{(\eta +1)^{3/2}}{(2\eta)^{3/2}}\left(\bar{\sigma}+\dfrac{\alpha c_h}{\mu}+\dfrac{c_f}{m^2}\pm \dfrac{1}{\mu l}\sqrt{\dfrac{1-\eta}{1+\eta}}\right)(r_{+}-r_{-}),
\end{equation}
where
\begin{equation}
c_{R/L}=\dfrac{3 l}{2G}\left(\bar{\sigma}+\dfrac{\alpha c_h}{\mu}+\dfrac{c_f}{m^2}\mp \dfrac{1}{\mu l}\right),
\end{equation}
and 
\begin{equation}
T_{R/L}=\dfrac{\sqrt{2(\eta+1)(b^2 l^2+4\mathcal{M})}}{8\pi l\left(1\pm \sqrt{\dfrac{1-\eta}{1+\eta}}\right)}=\dfrac{1}{8\pi l^2}\sqrt{\dfrac{2(\eta +1)}{\eta}}\dfrac{(r_{+}-r_{-})}{\left(1\pm \sqrt{\dfrac{1-\eta}{1+\eta}}\right)}.
\end{equation}
From (\ref{eqer1})$-$(\ref{eqentropy}), the first law of thermodynamics  
\begin{equation}
dM=T dS+\Omega dJ,
\end{equation}
is satisfied if
\begin{equation}
m^2=\dfrac{-\mu l^2 c_f (3\eta^3+\eta^2-3\eta-1)}{(4c_h \eta^2 l^2 \mu-4c_h \eta l^2 \mu+3\eta^2+2\eta-1)a-(1+\eta)l^2\left((\alpha c_h-3\mu \sigma)\eta^2+(\alpha c_h+\mu \sigma)(2\eta+1)\right)}.
\end{equation}
For GMG ($\alpha=0$), the thermodynamical quantities for rotating BTZ ($b=0$) from (\ref{eqer1})$-$(\ref{eqang1}) become
\begin{align}
E=&\dfrac{1}{8G}\left(\sigma +\dfrac{1}{2m^2 l^2}\right)\mathcal{M}+\dfrac{\mathcal{M}a}{4\mu G l^2},\\
J=&\dfrac{1}{8G}\left(\sigma +\dfrac{1}{2m^2 l^2}\right)\mathcal{M}a+\dfrac{\mathcal{M}}{4\mu G},
\end{align}
which are the same as the thermodynamical quantities for BTZ in \cite{Bergshoeff:2019rdb}.
\subsection{NEAR HORIZON SYMMETRIES}
By looking at the near horizon of the rotating black hole one can find
\begin{align}\label{eqq60}
\kappa=&\dfrac{\sqrt{2(\eta +1)(l^2 b^2+4\mathcal{M})}}{4 l},\hspace{0.5cm}\theta=\dfrac{\sqrt{2(1-\eta)(b^2 l^2+4\mathcal{M})}}{2},\hspace{0.5cm}\nonumber\\
&\gamma^2=\dfrac{l^2}{8\eta}(1+\eta)\left(-b^2 l^2 (1+\eta)-2 b l\sqrt{\eta(b^2 l^2+4\mathcal{M})}+4\mathcal{M}\right)
\end{align}
by using (\ref{eqq60}) and assuming the parameters independent of $\phi$, the charge $Q$ yields
\begin{equation}
Q=-\dfrac{1}{8G}\left[-\left(\left(\bar{\sigma}+\dfrac{\alpha c_h}{\mu}+\dfrac{c_f}{m^2}\right)\gamma +\dfrac{\theta}{4\mu}\right)\left(\left(\theta+\dfrac{\alpha c_h \gamma \theta}{2\mu}+\dfrac{\gamma^2}{2 l^2}\right)L+2\kappa P\right)\right].
\end{equation}
For the rotating solution, equations (\ref{eqqtau}) and (\ref{eqy}) become
\begin{equation}\label{eqtau2}
8 G \mathcal{T}_{n}=-\dfrac{\kappa}{2\mu l}\left[4\mu l^2 \left(\bar{\sigma}+\dfrac{\alpha c_h}{\mu}+\dfrac{c_f}{m^2}\right)\sqrt{\dfrac{(1+\eta)(\mathcal{M}+b r_{-})}{2\eta}}+\sqrt{\dfrac{1-\eta}{2\eta}}(r_{+}-r_{-})\right]\delta_{n,0}
\end{equation}
\begin{align}
8G \mathcal{Y}_{n}=&[\left(\bar{\sigma}+\dfrac{\alpha c_h}{\mu}+\dfrac{c_f}{m^2}\right)\sqrt{\dfrac{(1-\eta^2)(\mathcal{M}+br_{-})}{4\eta^2}}\left(r_{+}-r_{-}\right)+\dfrac{(1-\eta)}{8\mu l^2 \eta}\left(r_{+}-r_{-}\right)^2\nonumber\\
&+\dfrac{1+\eta}{2\mu \eta}\left(\mathcal{M}+b r_{-}\right)]\delta_{n,0}.
\end{align}
From equation (\ref{eqtau2}), one can obtain entropy
\begin{equation}
S=-\dfrac{\pi}{2\mu l}\left[4\mu l^2 \left(\bar{\sigma}+\dfrac{\alpha c_h}{\mu}+\dfrac{c_f}{m^2}\right)\sqrt{\dfrac{(1+\eta)(\mathcal{M}+b r_{-})}{2\eta}}+\sqrt{\dfrac{1-\eta}{2\eta}}(r_{+}-r_{-})\right]
\end{equation}
and $\mathcal{Y}_{0}=J$ is the angular momentum of the rotating black hole.
\begin{align}
&i[\mathcal{T}_{m},\mathcal{T}_{n}]=-\dfrac{\kappa^2 n}{8\mu}\delta_{m+n,0},\hspace{0.5cm}i[\mathcal{Y}_{m},\mathcal{Y}_{n}]=(m-n)\mathcal{Y}_{m+n},\nonumber\\
&i[\mathcal{Y}_{m},\mathcal{T}_{n}]=-n\mathcal{T}_{m+n}+\dfrac{\kappa n}{32 \mu l}\left[\sqrt{\dfrac{1-\eta}{2\eta}}(r_{+}-r_{-})+2\alpha c_h l^2 \sqrt{\dfrac{(1+\eta)(\mathcal{M}+b r_{-})}{2\eta}}\right]\delta_{m+n,0}
\end{align}
\section{Conclusion}
We considered static and rotating black holes in AdS with softly decaying hair. This metric, under the condition (\ref{eqsp}), is a solution for the GMMG theory. We have obtained the mass, angular momentum and entropy of the black hole. As can be seen the charges in addition to the standard mass parameter ($\mathcal{M}$), also depends on the gravitational hair ($b$). By using of available left and right central charges and computation the left and right temperatures, we have obtained the left and right entropy for the black hole and checked the first law of thermodynamics. We have obtained the asymptotic symmetries of the solution by considering AdS boundary condition (\ref{eqcond}). As can be seen in addition to local conformal symmetry, there is an extra symmetry which corresponds to the vanishing Noether charges and, therefore, turn out to be pure gauge. Then, by going to the Gaussian null coordinates, we looked at the near horizon symmetries. The near horizon Killing vectors are given by Eq. (\ref{horizonkilling}) and in Fourier modes obey the algebra (\ref{furiealgebra}). We have obtained the algebra of the conserved charges in Fourier modes. We have seen that the zero mode of $\mathcal{T}_{n}$ and $\mathcal{Y}_{n}$ reproduct entropy and angular momentum of black hole. Finally, we have applied the same analyses to the rotating black hole.

\section*{Acknowledgements} We thank Gaston Giribet for enlightening com-
ments. J. O. thanks the support of proyecto FONDECYT REGULAR 1221504 and 1210635 and Proyecto de Cooperación Internacional
2019/13231-7 FAPESP/ANID. After this work was accepted, a tragic
event led us to mourn the loss of Prof. M.R. Setare. May the publication
of this work, an idea proposed by him, contribute as a memory, which
will be forever with us. M. R. S and S. N. S acknowledge the support
of Kurdistan University.

\appendix
\section{Field equations}\label{app0}
The equations of motion of the Lagrangian of GMMG is obtained by making variation with
respect to the fields $e$, $\Omega=\omega-\alpha h$, $h$ and $f$ as follows \cite{Setare:2017mry}
\begin{small}
\begin{align}
E_e=&-\sigma R(\Omega)+\dfrac{{\Lambda}_0}{2}e\times e+(1+\sigma \alpha)Dh-\dfrac{\alpha}{2}(1+\sigma \alpha)h\times h-\dfrac{1}{2 m^2}f\times f=0\nonumber\\
E_\Omega =&-\sigma T(\Omega)+(1+\sigma \alpha)e\times h+\dfrac{1}{2\mu}\left(2R(\Omega)+\alpha^2 h\times h-2\alpha Dh\right)+\dfrac{1}{m^2}\left(Df-\alpha f\times h\right)=0\nonumber\\
E_h=&(1+\sigma \alpha)T(\Omega)-\alpha (1+\sigma \alpha)e\times h+\dfrac{1}{2\mu}\left(2\alpha^2 Dh-\alpha^3 h\times h-2\alpha R(\Omega)\right)
+\dfrac{1}{m^2}\left(-\alpha Df+\alpha^2 f\times h\right)=0\nonumber\\
E_f=&R(\Omega)-\alpha Dh+\dfrac{\alpha^2}{2}h\times h+e\times f=0
\end{align}
\end{small}
Here the covariant derivative $D(\Omega)A$ is defined by $D(\Omega)A = dA + \Omega \times A$ and its curvature $R(\Omega)=d\Omega+1/2\Omega\times \Omega$. Using the equations of motion for $\Omega$ and $h$ one can see that $\Omega$ is a torsion free connection,
 \begin{equation}\label{eqtor}
 T(\Omega)=0,
 \end{equation}
 which lead to the following set of equations
\begin{align}
&-\sigma R(\Omega)+\dfrac{\Lambda_0}{2}e\times e+(1+\sigma \alpha)Dh-\dfrac{\alpha}{2}(1+\sigma \alpha)h\times h-\dfrac{1}{2 m^2}f\times f=0,\label{eqeom} \\
&\mu(1+\sigma \alpha)e\times h-e\times f-\dfrac{\mu}{m^2}(Df-\alpha f\times h)=0,\label{eqeom1}\\
&R(\Omega)-\alpha Dh+\dfrac{\alpha^2}{2}h\times h+e\times f=0\label{eqeom2}.
\end{align}
These equations of motion admit a maximally symmetric AdS vacuum solution with cosmological constant $\Lambda$. In order to prove that, we choose the auxiliary fields to be proportional to dreibein
 \begin{equation}\label{eqfh}
 h=c_h e,\;\;\;\; f=c_f e,
 \end{equation}
and since for a constant curvature vacuum of curvature $\Lambda$ one has 
 \begin{equation}\label{eqR}
 R(\Omega)=\dfrac{1}{2}\Lambda e\times e=-\dfrac{1}{2l^2} e\times e
 \end{equation}
 the system (\ref{eqeom})-(\ref{eqeom2}) in this case reduces to the following algebraic system
 \begin{align}\label{eqcc}
& \dfrac{\sigma}{l^2}-\alpha(1+\alpha\sigma)c_{h}^2+\Lambda_{0}-\dfrac{c_{f}^{2}}{m^2}=0,\\
 &\dfrac{1}{\mu l^2}-2(1+\alpha\sigma)c_{h}-\dfrac{2\alpha}{m^2}c_{f}c_{h}-\dfrac{\alpha^2}{\mu}c_{h}^2=0,\\
 &c_{f}-\mu(1+\alpha\sigma)c_{h}-\dfrac{\mu\alpha}{m^2}c_{f}c_{h}=0 .
 \end{align}
By solving above equations, one can obtain $\Lambda_{0}$, $c_{h}$ and $c_{f}$ in terms of the couplings $\sigma,\mu,\alpha$ and $m^2$ as follows: 
 \begin{equation}
c_{f}=-\dfrac{3m^{2}}{\alpha}+\dfrac{\sqrt{3}m}{\mu \alpha \mathcal{C}^{\frac{1}{6}}}\left(\sqrt{\mathcal{E}}\pm\dfrac{\sqrt{\mathcal{D}}}{\mathcal{E}^{\frac{1}{4}}}\right),
\end{equation}
\begin{equation}
c_{h}=\dfrac{m^{2}\left(-3m\mu \mathcal{C}^{\frac{1}{6}}\mathcal{E}^{\frac{1}{4}}+\sqrt{3}\mathcal{E}^{\frac{3}{4}}\pm\sqrt{3\mathcal{D}}\right)}{\mu \alpha\left(6\mu \alpha m \sigma \mathcal{C}^{\frac{1}{6}}\mathcal{E}^{\frac{1}{4}}+3m \mu \mathcal{C}^{\frac{1}{6}}\mathcal{E}^{\frac{1}{4}}+\sqrt{3}\mathcal{E}^{\frac{3}{4}}\pm\sqrt{3\mathcal{D}}\right)},
\end{equation}
which explicit form of $\mathcal{C}, \mathcal{D}$ and $\mathcal{E}$ provided as follows

\begin{align}
\mathcal{B}&=27\mu^{6}(\sigma^{2}m^2+\Lambda_{0})\left(m^{2}(\sigma^{2}\alpha^2 +\dfrac{4}{3}\sigma \alpha+\dfrac{1}{3})-\dfrac{\Lambda_{0}\alpha^2}{3}\right)^{3}+\alpha^{2}m^{8}\mu^{2}(\sigma^{2}m^{4}(\sigma \alpha+1)^{2}+\nonumber\\
&m^{2}(3\Lambda_{0}-6\sigma \alpha \Lambda_{0}-10\sigma^{2}\alpha^{2}\Lambda_{0}))+\Lambda_{0}\alpha^{3}m^{12}-9\alpha m^{10}\mu^{4}\sigma^{2}(1+\sigma\alpha)^{2}(\sigma^{2}\alpha^{2}+\dfrac{2}{3}\sigma\alpha-\dfrac{2}{9})\nonumber\\
&3\alpha m^4 \mu^4\left(11\Lambda_{0}m^{4}(1+\sigma \alpha)(\sigma^{3}\alpha^3+\frac{23}{33}\sigma^{2}\alpha^{2}+\frac{1}{11}\sigma \alpha+\frac{1}{11})+\dfrac{29\alpha^2 \Lambda_{0}^{2}m^2}{3}(\sigma^{2}\alpha^{2}+\frac{48}{29}\sigma\alpha+\dfrac{21}{29})+\Lambda_{0}^{3}\alpha^{4}\right)\nonumber\\
\mathcal{C}&=6m^3\mu^2 \sqrt{3\mathcal{B}}\alpha^{\frac{3}{2}}(1+\sigma \alpha)+27\mu^{6}\left(m^{2}(\sigma^{2}\alpha^2+\frac{4}{3}\sigma \alpha+\frac{1}{3})-\frac{\Lambda_{0}\alpha^{2}}{3}\right)^{3}+\mu^{2}(3\alpha^{2}m^{10}(1-2\sigma\alpha-3\sigma^{2}\alpha)\nonumber\\
&-3\Lambda_{0}\alpha^{4}m^{8})+27\alpha m^{4}\mu^{4}\left(m^{4}(\sigma\alpha+1)^2(\sigma^2\alpha^2+\frac{1}{9})+2\alpha^2\Lambda_{0}m^2(\sigma\alpha+1)(\sigma\alpha+\frac{8}{9})+\dfrac{\Lambda_{0}^{2}\alpha^{4}}{9}\right)+m^{12}\alpha^{3}\nonumber\\
\mathcal{E}&=\mathcal{C}^{\frac{1}{3}}\left(m^2 \mu^2+2\alpha m^2(2\mu^2 \sigma-m^2)+6\mu^2\alpha^{2}(\sigma^{2}m^2+\dfrac{\Lambda_{0}}{3})\right)+\mathcal{C}^{\frac{2}{3}}+\mu^4\alpha^4(\Lambda_{0}^{2}-6\sigma^{2}m^2\Lambda_{0}+9\sigma^{4}m^{4})\nonumber\\
&+2\alpha^3 \mu^2 m^2\left(4\mu^2 \sigma(-\Lambda_{0}+3\sigma^{2}m^2)-m^{2}\left(\Lambda_{0}+3\sigma^{2}m^{2}\right)\right)+\alpha^{2}m^{2}(m^{4}-4\sigma m^{4}\mu^{2}+2\mu^{4}(11\sigma^{2}m^{2}-\Lambda_{0}))\nonumber\\
&+2\alpha m^{4}\mu^{2}\left(m^{2}+4\sigma \mu^2\right)+m^{4}\mu^{4}\nonumber\\
\mathcal{D}&=\mathcal{K}\sqrt{\mathcal{L}}+6\alpha m \mu \sqrt{3\mathcal{C}}\left(m^{4}+(\alpha \mu^{2}+2\sigma \mu^{2}\alpha^{2})(\Lambda_{0}+\sigma^{2}m^{2})\right)\nonumber\\
\mathcal{K}&=\mathcal{C}^{\frac{1}{3}}\left(12\mu^2\alpha^2(\sigma^2 m^2+\frac{\Lambda_{0}}{3})+4\alpha m^{2}(2\sigma \mu^2-m^2)+2m^{2}\mu^{2}\right)-\mathcal{C}^{\frac{2}{3}}+\mu^4 \alpha^4(-\Lambda_{0}^{2}-9\sigma^{4}m^{4}+6\sigma^{2}m^{2}\Lambda_{0})\nonumber\\
&\alpha^{3}\left(\mu^{4}(8m^{2}\sigma\Lambda_{0}-24\sigma^{3}m^{4})+\mu^{2}(2\Lambda_{0}m^{4}+6\sigma^{2}m^{6})\right)-2\alpha \mu^{2}m^{4}\left(m^{2}+4\sigma \mu^{2}\right)-\mu^{4}m^{4}+\nonumber\\
&\alpha^{2}\left(2\sigma m^{6}\mu^{2}-m^{8}+\mu^{4}(2m^{2}\Lambda_{0}-22\sigma^{2}m^{4})\right)\nonumber\\
\mathcal{L}&=\mathcal{C}^{\frac{1}{3}}\left(6\mu^{2}\alpha^{2}(\sigma^{2}m^{2}+\frac{\Lambda_{0}}{3})+2\alpha m^{2}(2\sigma \mu^{2}-m^{2})+m^{2}\mu^{2}\right)+\mathcal{C}^{\frac{2}{3}}+\mu^{4}\alpha^{4}\left(\Lambda_{0}^{2}-6\sigma^{2}m^{2}\Lambda_{0}+9\sigma^{4}m^{4}\right)\nonumber\\
&+\alpha^{3}\left(\mu^{4}(24\sigma^{3}m^{4}-8\sigma\Lambda_{0}m^{2})-\mu^{2}(2\Lambda_{0}m^{4}+6\sigma^{2}m^{6})\right)+\alpha^{2}\left(m^{8}-4\sigma m^{6}\mu^{2}+\mu^{4}(22\sigma^{2}m^{4}-2m^{2}\Lambda_{0})\right)\nonumber\\
&+2\alpha \mu^{2}m^{4}\left(m^{2}+4\sigma \mu^{2}\right)+\mu^{4}m^{4}.
\end{align}

\section{The mass excitation}\label{app01}
{Here we want to obtain the mass of gravitons. Denoting the vacuum solution by $\bar{e},\bar{\Omega}, \bar{h}$ and $\bar{f}$ then a general perturbation may be written as 
\begin{equation}
e=\bar{e}+k,\;\;\Omega=\bar{\Omega}+v,\;\; h=c_h (\bar{e}+k)+p,\;\;f=c_f (\bar{e}+k)+q, 
\end{equation}
 where $k, v, p$ and $q$ are small perturbations of $e, \Omega, h$ and $f$ respectively. Plugging this anstatz into the equations of motion given by (\ref{eqtor}), (\ref{eqeom}) and using (\ref{eqcc}) one arrives at
\begin{align}
&\bar{D}k+\bar{e}\times v=0 \label{eqkk}\\
&\bar{D}v-c_f\left(1+\sigma \alpha+\dfrac{\alpha}{2 m^2}\right)\bar{e}\times q-\Lambda \bar{e}\times k=0\label{eqvv}\\
&\bar{D}q+\left(\dfrac{m^2}{\mu}c_f-\alpha c_h\right)\bar{e}\times q+\left(m^2(1+\sigma\alpha)-\alpha c_h\right)\bar{e}\times p=0\label{eqqq}\\
&\bar{D}p-c_f\left(\sigma+\dfrac{1}{2m^2}\right)\bar{e}\times q-(\alpha c_h) \bar{e}\times p=0\label{eqpp}
\end{align}
After diagonalize the linear equations about an AdS vacuum, we set
%\begin{align}
%q&=...\\
%p&=...\\
%k&=....\\
%v&=....
%\end{align}
\begin{align}
&\bar{D}\phi_{+}+\Xi_{+}\bar{e}\times \phi_{+}=0,\\
&\bar{D}\phi_{-}+\Xi_{-}\bar{e}\times \phi_{-}=0,\\
&\bar{D}f_{+}+l^{-1}\bar{e}\times f_{+}=0,\\
&\bar{D}f_{-}-l^{-1}\bar{e}\times f_{-}=0,
\end{align}
where
\begin{equation}
\Xi_{\pm}=-\dfrac{s {m}^2}{2{\mu}}\pm\sqrt{\dfrac{1}{2l^2}+\bar{\sigma}^2s {m}^{2}+\dfrac{{m}^4}{4{\mu}^2}-\dfrac{\bar{\sigma} s m^2 \gamma}{2\mu^2 l^2}}.
\end{equation}
and $\phi_{+}, \phi_{-},f_{+}, f_{-}$ are new basis.\\

\section{Conserved Charges}\label{app02}

 Here, using an extension of the ADT method for Chern$-$Simons$-$like theories of gravity \cite{Bergshoeff:2019rdb}, we will obtain the conserved charges the black hole solution \eqref{eqmetric} of GMMG. The dreibein components of the metric can be chosen as
\begin{equation}
e^{0}=\sqrt{f}dt,\;\;\;e^{1}=r d\phi,\;\;\; e^{2}=\dfrac{dr}{\sqrt{f}}\ ,
\end{equation}
where
\begin{equation}
f=\dfrac{r^2}{l^2}+b r-\mathcal{M}\ .
\end{equation}
The zero$-$torsion condition (\ref{eqtor}) determines the dual Lorentz connection one form
\begin{equation}
\omega^{0}=-\sqrt{f}d\phi,\;\;\;\; \omega^{1}=-f^{'}dt,\;\;\; \omega^{2}=0\ .
\end{equation}
where $f^{'}=\partial_{r}f$. 
Therefore, for the constant curvature background obtained by setting $\mathcal{M}=b=0$ on \eqref{eqmetric} one has
\begin{equation}\label{eqderb}
\bar{e}^{0}=\dfrac{r}{l}dt,\;\;\;\; \bar{e}^{1}=r d\phi,\;\;\;\; \bar{e}^2=\dfrac{l}{r}dr\ ,
\end{equation}
 and
 \begin{equation}\label{eqomega0}
 \bar{\omega}^0=-\dfrac{r}{l}d\phi,\;\;\; \bar{\omega}^{1}=-\dfrac{r}{l^2}dt,\;\;\;\; \bar{\omega}^2=0\ .
 \end{equation}
Consequently we define
\begin{align}
\Delta e^{0}=\left(\sqrt{f}-\dfrac{r}{l}\right)dt,\;\;\;\Delta e^{1}=0,\;\;\;\; \Delta e^{2}=\left(\dfrac{1}{\sqrt{f}}-\dfrac{l}{r}\right)dr
\end{align}
and
\begin{equation}
\Delta \omega^{0}=-\left(\sqrt{f}-\dfrac{r}{l}\right)d\phi,\;\;\;\Delta \omega^{1}=-\left(f^{'}-\dfrac{r}{l^2}\right)dt,\;\;\;\Delta \omega^{2}=0.
\end{equation}
The action principle of GMMG \eqref{eqlag} can be written in a CS$-$like form \cite{Setare:2014zea} 
\begin{equation}\label{410}
L=\frac{1}{2}\hat{g}_{rs}a^{r}.da^{s}+\frac{1}{6}\hat{f}_{rst}a^{r}.(a^{s}\times a^{t})
\end{equation}
provided we identify
\begin{eqnarray*}
\hat{g}_{\omega e}=-\sigma, \hspace{0.5cm} \hat{g}_{eh}=1, \hspace{0.5cm}\hat{g}_{f\omega}=\frac{-1}{m^2},\hspace{0.5cm} \hat{g}_{\omega\omega}=\frac{1}{\mu} \\
\hat{f}_{e\omega\omega}=-\sigma \hspace{0.5cm}\hat{f}_{e h\omega}=1 \hspace{0.5cm}\hat{f}_{eff}= \frac{-1}{m^2},\hspace{0.5cm}\hat{f}_{\omega\omega\omega}= \frac{1}{\mu}\\
\hat{f}_{\omega\omega f}=\frac{-1}{m^2},\hspace{0.5cm}\hat{f}_{eee}=\Lambda_0 \hspace{0.5cm}\hat{f}_{ehh}=\alpha.
\label{eqcsf}
\end{eqnarray*}
and require
\begin{equation}
\bar{h}=c_h \bar{e}, \;\;\;\; \Delta \bar{h}=c_h \Delta \bar{e},\;\;\; \bar{f}=c_f \bar{e},\;\;\; \Delta \bar{f}=c_f \Delta \bar{e}.
\end{equation}
For a CS$-$like gravity theory with a background $\bar{a}^{r}$, the conserved charges are given by \cite{Bergshoeff:2019rdb}
\begin{equation}\label{generalcharge}
Q(\xi)=\dfrac{1}{8\pi G}\oint \Delta a^{r}.i_{\xi}\bar{a}^{s}\hat{g}_{rs},
\end{equation}
where $\Delta a^{r}=a^{r}-\bar{a}^{r}$ and $i_{\xi}\bar{a}^{s}=\zeta^{s}$.
Application of this formula to our setup yields
\begin{small}
\begin{equation}\label{eqform}
8\pi G Q^{\mu}(\xi)=\oint_{\Sigma} \left[-\sigma \left(\Delta e.\bar{\omega}_{\mu}+\Delta \omega.\bar{e}_{\mu}\right)+\left(\Delta e.\bar{h}_{\mu}+\Delta h.\bar{e}_{\mu}\right)-\dfrac{1}{m^2}\left(\Delta f.\bar{\omega}_{\mu}+\Delta\omega.\bar{f}_{\mu}\right)+\dfrac{2}{\mu}\Delta \omega.\bar{\omega}_{\mu}\right]\xi^{\mu}\ .
\end{equation}
\end{small}
Now, considering the Killing vector $\partial_t$ of the black hole vacuum background \eqref{eqmetric}, we find that
\begin{small}
\begin{align}\label{eqmasssta}
&E=Q(\partial_t)=\dfrac{1}{8\pi G}\oint \left[\sigma \Delta \omega^{0}\bar{e}^{0}_{t}-\Delta e^{0}\bar{h}^{0}_{t}-\Delta h^{0}\bar{e}^{0}_{t}+\dfrac{1}{m^2}\Delta \omega^{0}\bar{f}^{0}_{t}+\dfrac{2}{\mu}\Delta \omega^{1}\bar{\omega}^{1}_{t}\right]=\nonumber\\
&\dfrac{1}{8\pi G}\left(\sigma+\dfrac{c_f}{m^2}\right)\int_{\infty}\left(\dfrac{r^2}{l^2}-\dfrac{r}{l}\sqrt{f}\right)d\phi=\dfrac{1}{8G}\left(\sigma+\dfrac{c_f}{m^2}\right)(4\mathcal{M}+b^2 l^2)=\dfrac{1}{8G l}\left(\sigma+\dfrac{c_f}{m^2}\right)(r_{+}-r_{-})^2\ .
\end{align}
\end{small}

\section*{Rotating solution}
From (\ref{eqrot}), the dreibein components are 
 \begin{equation}\label{eqder1}
 e^{0}=\sqrt{P}dt,\;\;\;e^{2}=\dfrac{dr}{\sqrt{F}},\;\;\;e^{1}=\sqrt{r^2+r_{0}^2}\left(N^{\phi}dt+d\phi\right)
 \end{equation}
 where $P=N^2 F$. The constant curvature background obtain with $\mathcal{M} = b = a=0$, so by using (\ref{eqderb}) and (\ref{eqder1}) we have
 \begin{align}
 \Delta e^{0}=\left(\sqrt{P}-\dfrac{r}{l}\right)dt,\;\; \Delta e^{2}=\left(\dfrac{1}{\sqrt{F}}-\dfrac{l}{r}\right)dr,\;\;\Delta e^{1}=\sqrt{r^2+r_{0}^2}N^{\phi}dt+\left(\sqrt{r^2+r_{0}^2}-r\right)d\phi.
 \end{align}
  The zero$-$torsion condition determines the dual Lorentz connection one form
 \begin{align}\label{eqome1}
 \omega^{0}&=-\left(\dfrac{\sqrt{(r^2+r_{0}^2)F}N^{\prime\phi}}{2}+r N^{\phi}\sqrt{\dfrac{F}{r^2+r_{0}^2}}\right)dt-r\sqrt{\dfrac{F}{r^2+r_{0}^2}}d\phi,\nonumber\\
 \omega^{1}&=\left(\dfrac{P^{\prime}}{2N}-\dfrac{(r^2+r_{0}^2)N^{\phi}N^{\prime \phi}}{2N}\right)dt-\dfrac{(r^2+r_{0}^2)N^{\prime \phi}}{2N}d\phi\nonumber\\
 \omega^{2}&=-\sqrt{\dfrac{r^2+r_{0}^2}{F}}\dfrac{N^{\prime \phi}}{2N}dr
 \end{align}
 therefore by using (\ref{eqomega0}) and (\ref{eqome1}) one gets
 \begin{align}
 \Delta \omega^{0}&=-\left(\dfrac{\sqrt{(r^2+r_{0}^2)F}N^{\prime\phi}}{2}+r N^{\phi}\sqrt{\dfrac{F}{r^2+r_{0}^2}}\right)dt-\left(r\sqrt{\dfrac{F}{r^2+r_{0}^2}}-\dfrac{r}{l}\right)d\phi,\nonumber\\
  \Delta \omega^{1}&=\left(\dfrac{P^{\prime}}{2N}-\dfrac{(r^2+r_{0}^2)N^{\phi}N^{\prime \phi}}{2N}+\dfrac{r}{l^2}\right)dt-\dfrac{(r^2+r_{0}^2)N^{\prime \phi}}{2N}d\phi,\nonumber\\
 \Delta \omega^{2}&= -\sqrt{\dfrac{r^2+r_{0}^2}{F}}\dfrac{N^{\prime \phi}}{2N}dr.
 \end{align}
 From (\ref*{eqcsf})$-$(\ref*{eqform}) and for the Killing vector $\partial_{t}$ one can obtain the mass as follows
\begin{align}\label{eqer}
E=&Q(\partial_t)=\dfrac{1}{8\pi G}\oint \left[\sigma (\Delta \omega^{0}\bar{e}^{0}_{t}-\Delta e^{1}\bar{\omega}^{1}_{t})-\Delta e^{0}\bar{h}^{0}_{t}-\Delta h^{0}\bar{e}^{0}_{t}+\dfrac{1}{m^2}(\Delta \omega^{0}\bar{f}^{0}_{t}-\Delta f^{1}\bar{\omega}^{1}_{t})+\dfrac{2}{\mu}\Delta \omega^{1}\bar{\omega}^{1}_{t}\right]\nonumber\\
&=\dfrac{1}{8\pi G l^2}\int_{\infty} \left\lbrace \left(\sigma+\dfrac{c_f}{m^2}\right) \dfrac{r(r^2+r_{0}^2-r l \sqrt{F})}{\sqrt{r^2+r_{0}^2}}+\dfrac{r(r^2+r_{0}^2)N^{\prime \phi}}{\mu N}\right\rbrace d\phi\nonumber\\
&=\dfrac{1}{32 G}\left(\sigma+\dfrac{c_f}{m^2}\right)\left[4\mathcal{M}+b^2 l^2-\dfrac{b^2 a^2}{2}\right]+\dfrac{a}{32 G \mu l^2}\left[8\mathcal{M}+b^2 l^2 (1-\eta)\right],
\end{align} 
 and for Killing vector $\partial_\phi$, one can obtain angular momentum as follows 
\begin{align}\label{eqang}
J=&Q(\partial_\phi)=\dfrac{1}{8\pi G}\oint \left[\sigma (\Delta e^{0}\bar{\omega}^{0}_{\phi}-\Delta \omega^{1}\bar{e}^{1}_{\phi})+\Delta e^{1}\bar{h}^{1}_{\phi}+\Delta h^{1}\bar{e}^{1}_{\phi}+\dfrac{1}{m^2}(\Delta f^{0}\bar{\omega}^{0}_{\phi}-\Delta \omega^{1}\bar{f}^{1}_{\phi})-\dfrac{2}{\mu}\Delta \omega^{0}\bar{\omega}^{0}_{\phi}\right]\nonumber\\
&=\dfrac{1}{8\pi G}\int_{\infty} \left\lbrace \left(\sigma+\dfrac{c_f}{m^2}\right) \left(\dfrac{r(r^2+r_{0}^2)N^{\prime \phi}}{N}\right)+2c_h \Delta e^{1}\bar{e}^{1}_{\phi}-\dfrac{2}{\mu}\left(\dfrac{r^2}{l}\sqrt{\dfrac{F}{r^2+r_{0}^2}}-\dfrac{r^2}{l^2}\right)\right\rbrace d\phi\nonumber\\
&=\dfrac{1}{64 G}\left(\sigma+\dfrac{c_f}{m^2}\right)\left[8\mathcal{M}+b^2 l^2(1-\eta)\right]a+\dfrac{1}{32 G\mu}\left[(4\mathcal{M}+b^2 l^2) (1+\eta)-\dfrac{b^2 a^2}{2}\right]+\nonumber\\
&\dfrac{c_h l^2}{32 G}\left[(4\mathcal{M}+b^2 l^2)(1-\eta)-\dfrac{b^2 a^2}{2}\right].
\end{align}

\end{document}